\documentclass[11pt]{article}
\usepackage[latin9]{inputenc}
\usepackage[a4paper]{geometry}
\usepackage{color}
\usepackage{textcomp}
\usepackage{amsmath}
\usepackage{amssymb}
\usepackage{graphicx}
\usepackage{esint}
\usepackage[unicode=true,
 bookmarks=false,
 breaklinks=false,pdfborder={0 0 1},backref=false,colorlinks=false]
 {hyperref}

\makeatletter
\numberwithin{equation}{section}
\newcommand{\lyxaddress}[1]{
\par {\raggedright #1
\vspace{1.4em}
\noindent\par}
}

\newcommand{\bea}{\begin{eqnarray}}
\newcommand{\eea}{\end{eqnarray}}
\renewcommand\[{\begin{equation}}
\renewcommand\]{\end{equation}}

\makeatother

\begin{document}

\title{Entanglement Entropy from Corner Transfer Matrix in Forrester Baxter
non-unitary RSOS models}

\author{Davide Bianchini$^{a}$ and Francesco Ravanini$^{b,c}$}

\date{September 2015}

\maketitle

\lyxaddress{\begin{center}
$^{a}$Dept. of Mathematics, City University London, Northampton
Square, EC1V 0HB, London, UK\\
$^{b}$Dept. of Physics and Astronomy, University of Bologna, Via
Irnerio 46, 40126 Bologna, Italy\\
$^{c}$I.N.F.N. - Sezione di Bologna, Via Irnerio 46, 40126 Bologna,
Italy
\par\end{center}}
\begin{abstract}
Using a Corner Transfer Matrix approach, we compute the bipartite
entanglement Rényi entropy in the off-critical perturbations of non-unitary
conformal minimal models realised by lattice spin chains Hamiltonians
related to the Forrester Baxter RSOS models \cite{NostroPearce} in
regime III. This allows to show on a set of explicit examples that
the Rényi entropies for non-unitary theories rescale near criticality
as the logarithm of the correlation length with a coefficient proportional
to the effective central charge. This complements a similar result,
recently established for the size rescaling at the critical point
\cite{NostroLondra}, showing the expected agreement of the two behaviours.
We also compute the first subleading \emph{unusual correction} to
the scaling behaviour, showing that it is expressible in terms of
expansions of various fractional powers of the correlation length,
related to the differences $\Delta-\Delta_{\min}$ between the conformal
dimensions of fields in the theory and the minimal conformal dimension.
Finally, a few observations on the limit leading to the off-critical
logarithmic minimal models of Pearce and Seaton \cite{PearceSeaton}
are put forward.
\end{abstract}
Email: Davide.Bianchini.1@city.ac.uk, francesco.ravanini@bo.infn.it

\global\long\def\Bell{\boldsymbol{\ell}}
\global\long\def\ssum#1#2{\sideset{}{^{*}}\sum_{{#1}}^{{#2}}}
\global\long\def\ceff{c_{\mathrm{eff}}}
\global\long\def\eqdef{\overset{\mathrm{{\scriptstyle def}}}{=}}
\global\long\def\Bsigma{\boldsymbol{\sigma}}

\newpage{}

\section*{Introduction}

Entanglement is a very specific and intriguing feature of quantum
systems \cite{Schrodinger}. It describes truly quantum correlations
between parts of a system. Besides the questions of principal nature
that it arises in the interpretation and behaviour of quantum mechanics
\cite{EPR,Bell}, it finds very interesting and promising applications
in quantum information theory and quantum computing \cite{Bennet},
in condensed matter physics \cite{Amico,Latorre}, as well as in the
physics of black holes \cite{BlackHoles} and has even risen some
interest in biological systems \cite{Biology}.

Many proposals have been formulated to quantify it (see for example
\cite{EMeasures} and references therein). In this paper we focus
on bipartite systems, composed of a subsystem $A$ and its complement
$\bar{A}$. In such case, the most convenient measure of entanglement
is the Von Neumann Entropy restricted to $A$, the so-called \emph{Entanglement
Entropy} \cite{Bennet}. It is used as a measure of entanglement for
pure quantum states. The behaviour of this quantity has been deeply
studied in a wide range of systems from analytical, numerical and
experimental points of view. For example, interesting experimental
protocols in many body systems have been recently proposed \cite{EEExp,EEExp2}.
It has been evaluated in different dimensions and in different regimes,
but it is in $1+1$ dimensions that it shows its most amazing mathematical
properties. In particular, in the case of Integrable Models, it reflects
many mathematical features that otherwise would be difficult to probe.
For example, the study of Entanglement Entropy in critical one-dimensional
quantum systems is one of the most powerful techniques for the identification
of the central charge $c$ of the conformal field theory describing
the low energy excitations of the model. This evaluation can be performed
both analytically and numerically (e.g. DMRG \cite{DMRG}) and usually
requires less information than the study of the scaling of the ground
state energy for the identification of the universality class.

In the last decade there has been a particular focus on the evaluation
of Entanglement Entropy in Integrable models, in particular in Conformal
Field Theory (CFT) \cite{HLW,CardyCalabrese}, in their massive perturbations
\cite{EEQFT,Doyon2009} and in lattice spin chains \cite{CTMPeschel,Franchini2008,EEXYZ,Ercolessi2011,Ercolessi2012,Ercolessi2013}.
Recently also non unitary models have been taken into account, both
in the critical \cite{NostroLondra} and in the massive \cite{EENUQFT}
regimes.

At first sight, non unitary theories might be considered just as non
physical mathematical curiosities, but there are many very interesting
examples where they actually do play a physical role. For example,
it is known that a strongly interacting 2D electron gas in a magnetic
field produces edge modes described by CFT minimal models, in what
is known as Fractional Quantum Hall Effect (FQHE). In some cases it
has been shown that these edge modes are described by non unitary
CFT \cite{QH}. In \cite{QHNU}, for example, the minimal model $\mathcal{M}_{3,5}$
has been considered. The non-unitarity arises from the fact that in
this particular case the bulk is gapless (while in the ordinary case
it is gapped) and then the edge can dissipate into the bulk.

These excitations can be represented using generalised spin chains:
the critical Fibonacci Chain (the ``golden chain'' \cite{GoldenChain},
minimal model $\mathcal{M}_{4,5}$), which represents the FQHE with
filling $\nu=\frac{12}{5}$ and its non-unitary generalisation to
the Lee-Yang Model $\mathcal{M}_{2,5}$ \cite{Ardonne} are among
the most known chain representations. These critical chains all belong
to the wide class of $\mathcal{U}_{q}(sl(2))$ invariant spin chains
\cite{PasquierSaleur} in the universality class of conformal minimal
models $\mathcal{M}_{m,m'}$, unitary or not. They can be further
generalised in order to include their massive $\Phi_{13}$ perturbations.
The off-critical Hamiltonians of such chains have been recently obtained
\cite{NostroPearce} and they are related to the RSOS models \cite{ABF,FB},
through the usual connection between the time evolution operators
of 1D quantum spin chains and the row-to-row transfer matrix of 2D
classical lattice models.

The Hamiltonians of these chains are generically of the pseudo-hermitian
type, i.e. they are not hermitian but have real eignevalues. In recent
years they attracted a lot of interest in the description of many
physical phenomena, ranging from optical effects to non-equilibrium
systems, etc... Similarly, in Quantum Field Theory it has been widely
discussed \cite{WattsRSOS} how the scattering matrices of apparently
non-hermitian theories can be physically well defined. In particular,
using the definitions of \cite{WattsRSOS}, non-unitary theories can
be classified as \textit{real} or \textit{non real}; in both cases
the inner product is indefinite, but in real theories eigenenergies
are real and the eigenvalues of the S matrix are pure phases, while
in non real models the energies are not real and the S matrix has
eigenvalues which are not pure phases. Notice that if the S matrix
is a pure phase, there is a conservation of probability through the
scattering process; this conservation is a fundamental requirement
for having a real, well-defined quantum physical theory.

Our goal in this paper is to compute the bipartite Entanglement Entropy
of the ground state for spin chain hamiltonians of the type introduced
in \cite{NostroPearce} on an infinite one dimesional lattice, taking
as $A$ the negative semi-axis and as $\bar{A}$ the positive one.
For non-Hermitian hamiltonians this may sound ambiguous, as non Hermitian
operators share the same right and left eigenvalues but have different
right and left eigenvectors. However, one can argue \cite{NostroLondra}
using PT symmetry and chiral factorization of CFT that, at least at
criticality, the left and right ground states do coincide, thus leading
to a correct positive definition of the Entanglement Entropy for this
state.

This bipartite Entanglement Entropy calculation can be achieved by
considering the \emph{Corner Transfer Matrix} (CTM) introduced by
Rodney Baxter \cite{Baxter-book} in integrable statistical systems
on a 2D square lattice. This tool, originally used for the evaluation
of partition functions and one point functions of the 2D models, has
been extended to the evaluation of the Entanglement Entropy of the
corresponding 1D chains, following an approach developed in \cite{CTMPeschel}.

It is known that these chains, as well as their classical 2D counterparts,
the class of RSOS$_{r,s}$ models, show different regimes with different
physical behaviour. In the unitary case $s=1$, Andrews, Baxter and
Forrester (ABF) \cite{ABF} classified four regimes. There are two
second order phase transitions: one between regimes I and II and the
other between regimes III and IV. The identification of universality
classes at the transition points has been studied by Huse \cite{Huse}.
Approaching the III-IV transition the RSOS$_{r,1}$ models, for different
$r$, are described by the universality class of unitary CFT minimal
models $\mathcal{M}_{r-1,r}$. In the I-II phase transition, instead,
the approach to criticality is governed by the universality class
of $\mathbb{Z}_{L}$ parafermionic CFTs \cite{FatZam} with $L=r-2$.

In the non unitary cases $s\not=1$ explored in \cite{FB}, the classification
of regimes is more complicated. However, still there are regimes III
and IV (or X if $r$ odd and $s=\frac{r-1}{2}$) with a phase transition
between them identified \cite{Riggs,Nakanishi} with the universality
class of non-unitary minimal models $\mathcal{M}_{r-s,r}$. The identification
of the universality class of the regime I-II phase transition is still
an open problem in the $s\not=1$ case. We shall not address this
problem here and in the following we restrict our calculations to
regime III.

In the case of unitary ABF RSOS models, Franchini and De Luca \cite{EEABF}
have recently used the CTM technique to compute the Entanglement Entropy,
checking that in the phase transition III-IV it correctly gives the
results expected in the conformal unitary minimal models $\mathcal{M}_{r-1,r}$
while in the regime I-II phase transition, it gives those expected
for the $\mathbb{Z}_{r-2}$ parafermionic CFTs.

In this paper we extend this Entanglement Entropy calculation to the
regime III of $s\not=1$ RSOS models. The main goal of this investigation
is to test on a concrete example of non unitary integrable theories
where the calculations can be carried out exactly, the conjecture
that the Entanglement Entropy scales logarithmically in the correlation
length in a manner that parallels the scaling on a finite interval
at criticality (for details of what we mean here see subsect. \ref{sec:R=0000E9nyi-Entanglement-Entropies}).

The paper is organised as follows:
\begin{itemize}
\item In section \ref{sec:CFT-minimal-models} we recall a few notions on
minimal CFT models (unitary or not) and their characters, emphasising
their relation with modular forms.
\item Then in section \ref{sec:R=0000E9nyi-Entanglement-Entropies} we introduce
the basic concept and formulae of Rényi and Von Neumann Entanglement
Entropies. We discuss in particular their behaviour for finite size
of the subsystem at criticality and for infinte size but off-criticality.
We consider in both cases the dominating logarithmic term, but also
the form of the expected corrections.
\item In section \ref{RSOS} we describe the Forrester-Baxter (FB) RSOS
models whose continuum scaling limit is described by the perturbed
minimal model $\mathcal{M}_{m,m'}+t\Phi_{1,3}$ (where $t>0$ is a
perturbing parameter and $\Phi_{1,3}$ the least relevant operator). 
\item We briefly summarise in section \ref{sec:The-Corner-Transfer} the
Corner Transfer Matrix construction and review its use as a tool for
the evaluation of partition functions and Rényi Entropy. In particular,
specialising to the RSOS models, we emphasise the connection between
the blocks in the partition functions on multi-sheeted surfaces with
the conformal characters of the corresponding CFT at criticality.
\item Applying the CTM tool to the FB RSOS model, in section \ref{EERSOS}
we evaluate the Rényi Entropy for these theories. Our main results
are the confirmation of the presence of the \textit{effective central
charge} in the leading scaling of the Entropy \cite{NostroLondra}.
The difference from the unitary case where the usual central charge
appears is due to the fact that the physical ground state and the
conformal vacuum do not coincide in non unitary models.
\item \textcolor{black}{In section \ref{sec:Unusual-corrections} we evaluate
explicitly the most relevant corrections -- nowadays traditionally
called }\textcolor{black}{\emph{unusual corrections}}\textcolor{black}{{}
following \cite{Cardy-Calabrese-corr} -- to the dominating logarithmic
behaviour. They exhibit power law decays with exponents given by a
sort of }\textcolor{black}{\emph{effective dimensions}}\textcolor{black}{,
i.e. the difference between the conformal dimensions of some of the
relevant fields of the CFT universality class and the lowest (negative)
conformal dimension that always appears in non unitary CFTs.}
\item \textcolor{black}{In section \ref{sec:Log-CFT} we briefly comment
on the evaluation of Rényi Entropy in off-critical Logarithmic Minimal
Models, which can be seen as particular limit realisations of the
FB RSOS model. The observed absence of the double log behaviour is
related to the destruction of the Jordan block structures responsible
of the logarithmic behaviour as soon as the log-CFT are perturbed
off-criticality.}
\item Finally, in section \ref{sec:Conclusions-and-outlook} we trace our
conclusions and perspectives for future work.
\end{itemize}

\section{CFT minimal models and their characters\label{sec:CFT-minimal-models}}

We recall here some basic facts about conformal minimal models that
we need in the following. The minimal models $\mathcal{M}_{m,m'}$
are conformal theories whose Hilbert space is built up of two chiral
parts each one composed of a finite number of irreducible highest-weight
representations (HWR) of Virasoro algebra at a given value of its
central charge $c$. They are labelled by two coprime integers $m,m'$
such that $ $$2\leq m<m'$ and have central charge
\[
c=1-\frac{6(m-m')^{2}}{mm'}
\]
The (left) conformal families $[\Phi_{a,a'}]$ are labelled by two
integers $a,a'$ running on the domain $\mathcal{J}=\{(a,a')\,:\,1\leq a\leq m-1\,,\,1\leq a'\leq m'-1\}$.
The conformal dimensions of the primary states $|a,a'\rangle$ of
such families are given by
\[
\Delta_{a,a'}=\frac{(am'-a'm)^{2}-(m-m')^{2}}{4mm'}
\]
The $\mathbb{Z}_{2}$ symmetry $\Delta_{a,a'}=\Delta_{m-a,m'-a'}$
is present in the whole series of models.

The models are unitary for $m'=m+1$, non unitary otherwise. In the
non unitary case the state of lowest conformal dimension is not the
conformal vacuum $|0\rangle\equiv|\Phi_{1,1}\rangle$ but a different
state $|\mathrm{min}\rangle$ with negative conformal dimension (that
can be proven always to exist and be unique)
\[
\Delta_{\mathrm{min}}=\frac{1-(m-m')^{2}}{4mm'}
\]
Correspondingly, an effective central charge can be defined
\[
\ceff=c-24\Delta_{\min}=1-\frac{6}{mm'}
\]
Notice that while the central charge $c$ can be negative in non unitary
models, the effective one $\ceff$ is always positive.

The characters of the Virasoro HWRs can be written as
\[
\chi_{a,a'}(q)=\frac{q^{\Delta_{a,a'}-\frac{c}{24}}}{(q)_{\infty}}\sum_{k=-\infty}^{\infty}\left[q^{k(kmm'+am'\text{\textminus}a'm)}-q^{(km+a)(km'+a')}\right]
\]
where $q=e^{2\pi i\tau}$ and $(q)_{\infty}=\prod_{j=1}^{\infty}(1-q^{j})$.
They form a unitary representation of the modular group \cite{DiFrancesco}
$PSL(2,\mathbb{Z})$ whose generators are 
\begin{eqnarray*}
\hat{\mathcal{T}}\, & : & \,\tau\to\tau+1\\
\hat{\mathcal{S}}\, & : & \,\tau\to-\frac{1}{\tau}
\end{eqnarray*}
\begin{eqnarray}
\hat{\mathcal{T}}\chi_{a,a'}(q) & = & \chi_{a,a'}(qe^{2\pi i})=e^{2\pi i\left(\Delta_{a,a'}-\frac{c}{24}\right)}\chi_{a,a'}(q)\nonumber \\
\hat{\mathcal{S}}\chi_{a,a'}(q) & = & \chi_{a,a'}\left(\tilde{q}\right)=\sum_{(b,b')\in\mathcal{J}}\mathcal{S}_{a,a'}^{b,b'}\chi_{b,b'}(q)\label{eq:modular}
\end{eqnarray}
where $\tilde{q}=e^{-\frac{2\pi i}{\tau}}$. $\mathcal{S}$ is the
modular matrix 
\[
\mathcal{S}_{a,a'}^{b,b'}=2\sqrt{\frac{2}{mm'}}(-1)^{1+ab'+a'b}\sin\left(\pi\frac{m}{m'}ab\right)\sin\left(\pi\frac{m'}{m}a'b'\right)
\]
Introducing the elliptic function 
\begin{equation}
E(z,q)=\sum_{k\in\mathbb{Z}}(-1)^{k}q^{\frac{k(k-1)}{2}}z^{k}=\prod_{n=1}^{\infty}(1-q^{n-1}z)(1-q^{n}z^{-1})(1-q^{n})\label{eq:E}
\end{equation}
they can equivalently be written as
\begin{equation}
\chi_{a,a'}(q)=\frac{q^{\Delta_{a,a'}-\frac{c}{24}}}{(q)_{\infty}}\left\{ E\left(-q^{m'a-a'm+mm'},q{}^{2mm'}\right)-q^{aa'}E\left(-q^{m'a+a'm+mm'},q^{2mm'}\right)\right\} \label{eq:character-E}
\end{equation}
The function $E(z,q)$ is related to the first Jacobi theta function
\begin{equation}
\vartheta_{1}(u;p)=\sum_{n\in\mathbb{Z}}(-1)^{n-\frac{1}{2}}p^{\left(n+\frac{1}{2}\right)^{2}}e^{(2n+1)iu}=2p^{\frac{1}{4}}\sin u\prod_{n=1}^{\infty}(1-2p^{2n}\cos2u+p^{4n})(1-p^{2n})\label{eq:Jacobi1}
\end{equation}
by the relation 
\begin{equation}
E\left(e^{2iw},q\right)=iq^{-\frac{1}{8}}e^{iw}\vartheta_{1}(w;q^{\frac{1}{2}})\label{eq:E_theta1}
\end{equation}
The first Jacobi theta function, under a modular $\hat{\mathcal{S}}$
tranformation, behaves as
\begin{equation}
\vartheta_{1}(z,e^{-\frac{i\pi}{\tau}})=-i\sqrt{i\tau}e^{\frac{i\tau z^{2}}{\pi}}\vartheta_{1}(\tau z;e^{i\pi\tau})\label{eq:theta1_modular}
\end{equation}
These formulae will be useful later in sect. \ref{EERSOS} for the
computation of the Rényi Entropy. 

The minimal models can be perturbed off-criticality by picking up
combinations of their relevant fields, resulting in super-renormalizable
theories. In particular, the perturbation by the least relevant field
$\Phi_{1,3}$ preserves the integrability of the model off-criticality.
It is this integrable perturbation that represents the scaling limit
of the RSOS models in the off-critical vicinity of the phase transition
between regimes III and IV.

\section{Rényi Entanglement Entropies\label{sec:R=0000E9nyi-Entanglement-Entropies}}

Consider a bipartite quantum system whose Hilbert space is $\mathcal{H}=\mathcal{H}_{A}\otimes\mathcal{H}_{\bar{A}}$.
If the system is in a pure state $|\Psi\rangle$ (i.e. one with density
matrix $\rho=|\Psi\rangle\langle\Psi|$), the entanglement between
the two parts $A$ and its complement $\bar{A}$ can be described
by the family of Rényi Entanglement Entropies $S_{n}$ ($n\in\mathbb{R}_{>0}$)
\begin{equation}
S_{A}^{(n)}=\frac{1}{1-n}\log\,\mbox{tr}_{A}\rho_{A}^{n}\label{Renyi}
\end{equation}
where $\rho_{A}=\mbox{tr}_{\bar{A}}\rho$ is the reduced density matrix
of the subsystem $A$. The Von Neumann Entanglement Entropy $S=\lim\limits _{n\rightarrow1}S_{A}^{(n)}=-\mbox{tr}{}_{A}\left(\rho_{A}\log\rho_{A}\right)$
is a special case of the Rényi Entropies.

In \cite{HLW,CardyCalabrese} it has been shown that the Rényi Entropy
for a bipartite one-dimensional quantum system scales as
\begin{equation}
S_{A}^{(n)}\sim\frac{c}{6}\frac{n+1}{n}\log\frac{l}{\epsilon}\label{Renyi2}
\end{equation}
where $l$ is the size of the subsystem $A$, and $\epsilon$ is some
ultraviolet cut-off. This scaling holds when the size of the whole
system is much larger than the size $l$ of subsystem $A$ and when
the correlation length $\xi$ is much larger than $l$. This condition
on the correlation length is equivalent to assume that the system
is approaching criticality with a universality class identified by
a CFT with central charge $c$. In \cite{Cardy-Calabrese-corr} the
subleading contributions to this behaviour were estimated and look
like
\begin{equation}
S_{A}^{(n)}\sim\frac{c}{6}\frac{n+1}{n}\log\frac{l}{\epsilon}+a^{(n)}+b^{(n)}l^{-2x/n}+b^{\prime(n)}l^{-2x}+...\label{eq:Renyi-l}
\end{equation}
The constants $a^{(n)}$, $b^{(n)}$ and $b^{\prime(n)}$ are non-universal,
but the exponent $x$ is, and has to be identified with the scale
dimension $x=\Delta+\bar{\Delta}$ of some operator $\Phi$ of the
CFT universality class having the lowest conformal dimension among
those concurring to the corrections. Higher order corrections are
expected and they are due to other primary fields and to descendents.
For $n>1$ the $l^{-2x}$ term is subleading with respect to the $l^{-2x/n}$
one. Viceversa for $n<1$. Taking into account both terms is necessary
to ensure a smooth limit for $n\to1$.

On the other hand, if the correlation length $\xi$ is finite and
the size $l\rightarrow\infty$ we have \cite{Cardy-Peschel-Calabrese}
a similar expansion
\begin{equation}
S_{A}^{(n)}\sim\frac{c}{12}\frac{n+1}{n}\log\xi+A^{(n)}+B^{(n)}\xi^{-h/n}+B^{\prime(n)}\xi^{-h}+...\label{eq:Renyi-xi}
\end{equation}
which represents the entanglement between two infinite parts of an
infinite system. In this case the size $l$ of the subsystem $A$
has been replaced by the correlation length. This is due to the fact
that the leading term of the Entropy is given by the smallest physical
length which plays a role in the system. The factor of 2 difference
in the leading terms and in the corrections is simply explained by
the fact that the number of boundary points dividing $A$ from $\bar{A}$
is 2 in the first case and 1 in the second. The non-universal coefficients
of the corrections $A^{(n)}$, $B^{(n)}$and $B^{\prime(n)}$ in general
can be different from the $a^{(n)},b^{(n)},b^{\prime(n)}$ of eq.(\ref{eq:Renyi-l}).
Also, the universal exponent $h$ can differ from the $x$ above \cite{Ercolessi2012,EEABF}.

Recently \cite{NostroLondra} this evaluation has been extended in
order to take into account non unitary systems. The Hamiltonian operator
$\hat{H}$ describing a CFT system of Hilbert space $\mathcal{H}$
put on a cylinder of circumference $R$ is given by
\[
\hat{H}=\frac{2\pi}{R}\left(L_{0}+\bar{L}_{0}-\frac{c}{12}\right)
\]
where $L_{n},\bar{L}_{n}$ are Virasoro algebra generators and $c$
is their central charge. In unitary CFT the conformal vacuum $|0\rangle$
(defined as the unique state such that $L_{n}|0\rangle=0$, $\forall n\geq-1$)
is the physical ground state, i.e. $\forall|\psi\rangle\in\mathcal{H}:$
$\langle0|\hat{H}|0\rangle\leq\langle\psi|\hat{H}|\psi\rangle$).
Thus the ground state energy will scale as $E_{0}=-\frac{\pi c}{6R}$
\cite{Affleck,CBN}.

In non unitary CFT instead, there is at least one primary state $|\mathrm{min}\rangle$
with negative conformal weight $\Delta_{\mathrm{min}}$ ($\langle\mathrm{min}|L_{0}|\mathrm{min}\rangle=\Delta_{\mathrm{min}}<0$).
Obviously, this state and not the conformal vacuum $|0\rangle$ has
the lowest possible energy on the cylinder and is the true ground
state of the theory. For this reason the ground state energy scales
as $E_{\mathrm{min}}=-\frac{\pi\ceff}{6R}$, with the so called \emph{effective
central charge} $\ceff=c-24\Delta_{\mathrm{min}}\geq0$ instead of
the central charge $c$ \cite{ISZ}. The difference between the conformal
vacuum and the physical ground state plays an important role in the
definition and calculation of the Entanglement Entropy. For a detailed
discussion, see \cite{NostroLondra}. As a result, in non-unitary
CFT the Rényi Entropy scales as
\[
S_{A}^{(n)}\sim\frac{\ceff}{6}\frac{n+1}{n}\log\frac{l}{\epsilon}
\]
Although this result seems natural, its proof in a generic non unitary
CFT is far from trivial and can be acheived by a careful analysis
of how to correctly define twist operators in such case through an
orbifold approach \cite{NostroLondra}.

In the case of infinite subsystem, one is tempted to replace $\frac{l}{\epsilon}$
by $\xi$, the correlation length, as in the unitary cases. In this
paper our aim is to verify that this conjecture is correct in a concrete
lattice realization of an off-critical non-unitary model. We use the
RSOS non-unitary models of Forrester Baxter in regime III as lattice
regularizations of the perturbed CFT $\mathcal{M}_{m,m'}+t\Phi_{1,3}$
(where $t>0$ is a perturbing parameter).

In other words, we try in this paper to reproduce a formula of the
type
\begin{equation}
S_{A}^{(n)}\sim\frac{\ceff}{12}\frac{n+1}{n}\log\xi+A^{(n)}+B^{(n)}\xi^{-\frac{h}{n}}+B^{\prime(n)}\xi^{-h}\label{eq:EE-xi-expansion}
\end{equation}
calculating it from the exact expression for the Rényi Entanglement
Entropy obtainable by a Corner Transfer Matrix approach combined with
the evaluation of the correlation length $\xi$ on the lattice. We
expect that the exponent $h$ is affected by the non unitarity in
a way similar to the change from $c$ to $\ceff$ in the leading logarithmic
term.

\section{RSOS FB Models\label{RSOS}}

\begin{figure}
\centering{}\includegraphics[scale=0.4]{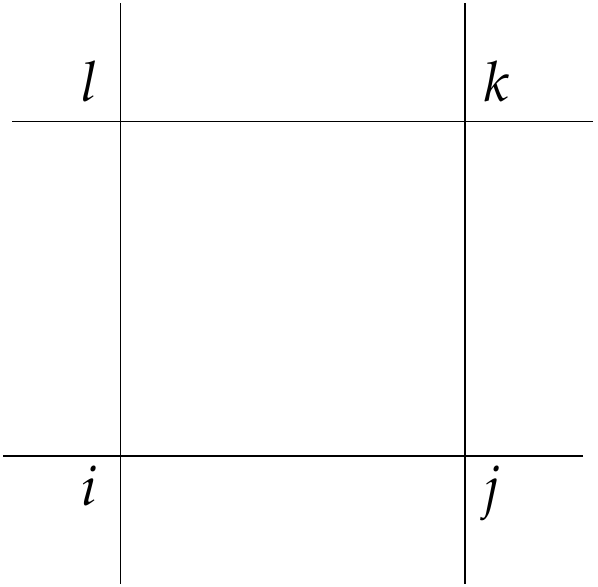}\caption{Square tile\label{fig:Square-tile}}
\end{figure}

The Restricted-Solid-On-Solid RSOS$_{r,s}$ model \cite{ABF,FB} is
defined on a square lattice, for each pair of coprime positive integers
$r,s$ such that $r>3$ and $1\leq s\leq r-1$. On each site $i\in\mathbb{Z}^{2}$
there is a variable, called \textit{local height}, $\ell_{i}$ that
takes values $\ell_{i}=1,2,\dots,r-1$. Local heights of two neighbouring
sites $i$ and $j$ are restricted to differ by one $|\ell_{i}-\ell_{j}|=1$.
Local heights can be thought as encoded on a $A_{r-1}$ Dynkin diagram.
Each node represents a possible value that $\ell_{i}$ can take and
it is linked to the possible values at neighbouring sites. This allows
to generalise the model, following Pasquier \cite{Pasquier}, to other
simply laced $A,D,E$ Dynkin diagrams. For the scope of the present
paper, however, we focus on the $A$ case only. The RSOS models belong
to the wide class of 2D classical lattice model of IRF (Interaction
Round a Face) type. In IRF models, the interaction is on nearest neighbour,
so that one can define local Boltzmann weights that depend on the
four sites $i,j,k,l$ around a tile (see fig. \ref{fig:Square-tile})

\[
W\left(\begin{array}{cc}
\ell_{l} & \ell_{k}\\
\ell_{i} & \ell_{j}
\end{array}\right)=e^{-\beta\varepsilon_{ij}^{lk}}
\]
where $\beta$ is the inverse temperature, $\varepsilon_{ij}^{lk}$
is the energy of the configuration of the four vertices.

For the RSOS models, the Boltzmann weights have been calculated from
Yang-Baxter equation, that must be satisfied by any integrable lattice
model, in \cite{ABF,FB}. They depend on a \emph{spectral parameter
$u$,} measuring the anisotropy of the lattice, on the \emph{crossing
parameter} 
\[
\lambda=\frac{s}{r}\pi
\]
ruling their behaviour when the lattice is rotated by $\frac{\pi}{2}$
\[
W\left(\left.\begin{array}{cc}
\ell_{l} & \ell_{k}\\
\ell_{i} & \ell_{j}
\end{array}\right|u\right)=\sqrt{\frac{s(\ell_{i}\lambda)s(\ell_{k}\lambda)}{s(\ell_{j}\lambda)s(\ell_{l}\lambda)}}W\left(\left.\begin{array}{cc}
\ell_{k} & \ell_{j}\\
\ell_{l} & \ell_{i}
\end{array}\right|\lambda-u\right)
\]
and on a \emph{temperature-like parameter} $p$, ($-1<p<1$), such
that $t=p^{2}$ is measuring the departure from criticality. The system
is critical for $p=0$. The Boltzmann weights turn out to be expressible
in terms of elliptic theta functions for which the parameter $p$
plays the rôle of the nome. Here and below $s(u)=\vartheta_{1}(u;p)$,
the first Jacobi theta function (\ref{eq:Jacobi1}). The non-zero
Boltzmann weights can be put in the following form (in the so called
symmetric gauge)
\begin{align}
W\left(\left.\begin{array}{cc}
\ell\pm1 & \ell\\
\ell & \ell\mp1
\end{array}\right|u\right) & =\frac{s(\lambda-u)}{s(\lambda)}\nonumber \\
W\left(\left.\begin{array}{cc}
\ell & \ell\pm1\\
\ell\mp1 & \ell
\end{array}\right|u\right) & =\frac{\sqrt{s((\ell\mp1)\lambda)s((\ell\pm1)\lambda)}}{s(\ell\lambda)}\frac{s(u)}{s(\lambda)}\label{eq:W}\\
W\left(\left.\begin{array}{cc}
\ell & \ell\pm1\\
\ell\pm1 & \ell
\end{array}\right|u\right) & =\frac{s(\ell\lambda\pm u)}{s(\ell\lambda)}\nonumber 
\end{align}
The periodicity and modular properties of the Boltzmann weights determine
the fundamental ranges of the parameters. The classification of possible
regimes of the models is quite complicated \cite{FB} but in this
paper we concentrate on regime III ($0<p<1$ and $0<u<\lambda$) only,
leaving the others for future investigation. 

For $s=1$ the model, often denoted simply RSOS$_{r}$, has been investigated
by Andrews, Baxter and Forrester in \cite{ABF}. The Rényi Entanglement
Entropy for the cooresponding spin chains has been computed in \cite{EEABF},
where it is shown that in the regime III approaching the critical
point $p=0$ it scales like (\ref{eq:Renyi-xi}) with the central
charge of unitary minimal models $\mathcal{M}_{r-1,r}$. In the regime
II transition, instead, it scales as the parafermions $\mathbb{Z}_{r-2}$.

\textcolor{black}{Here we are interested in the generalizations of
these results for other values of $s$. The spin chains Hamiltonians
corresponding to the RSOS$_{r,s}$ models have been recently written
in \cite{NostroPearce}. As the notation for them is quite complicated
and not relevant for the following, we do not write them here and
invite the interested reader to refer to the paper \cite{NostroPearce}
for an exhaustive presentation.}

\section{The Corner Transfer Matrix approach\label{sec:The-Corner-Transfer}}

For spin chains on an infinite 1D lattice, we consider the Renyi Entanglement
Entropy between two semi-infinite halves, the negative one conventionally
called $A$ while the positive is $\bar{A}$. For chains related to
a 2D classical lattice model of IRF (Interaction Round a Face) type,
an efficient method to compute the reduced density matrix $\rho_{A}$,
as proposed in \cite{CTMPeschel}, is to make use of the Corner Transfer
Matrix approach \cite{Baxter-book}. The FB RSOS models are of this
kind, therefore we adopt this approach here.

\begin{figure}
\centering{}\includegraphics[scale=0.5]{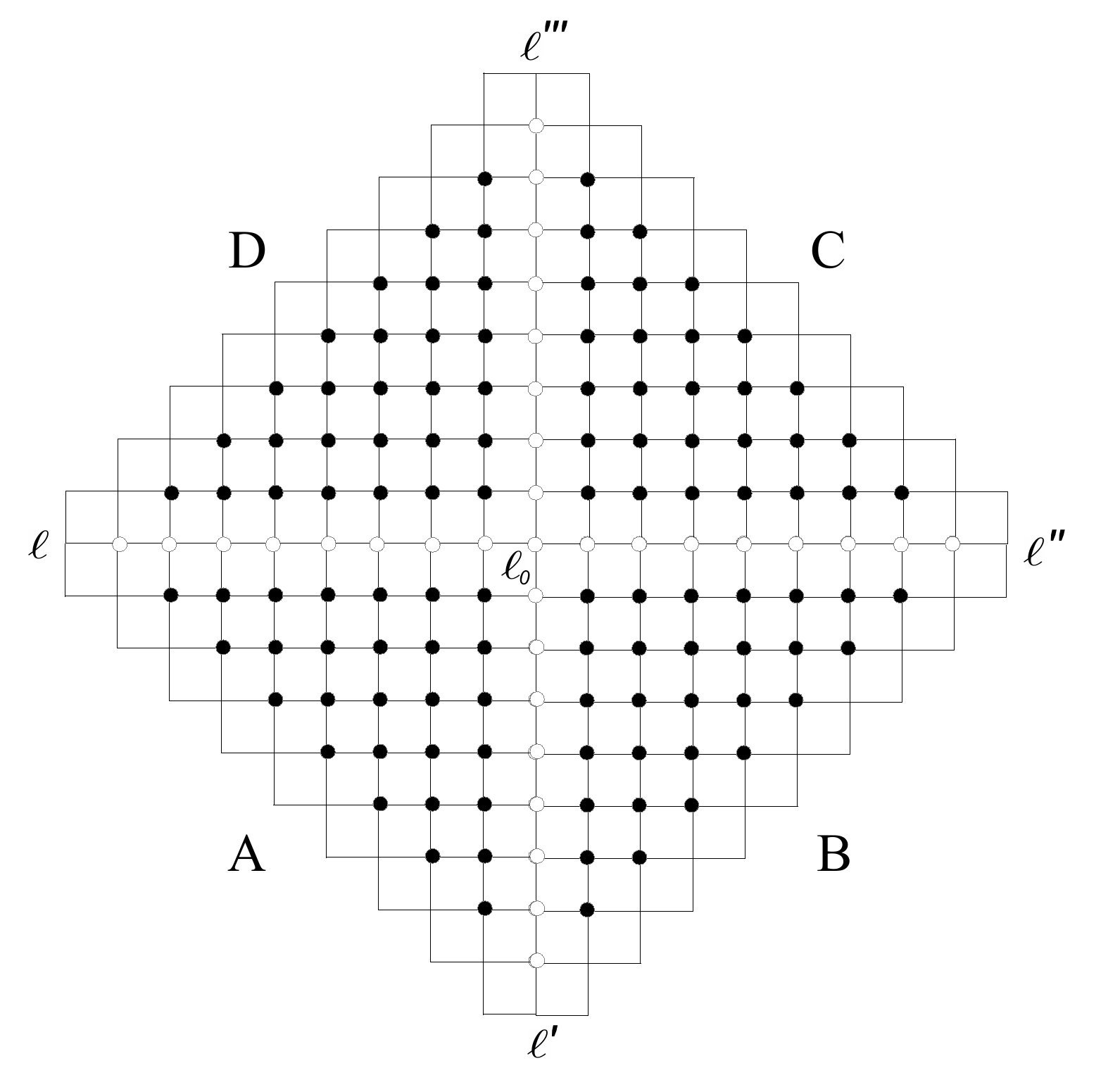}\caption{The 4 corner transfer matrix operators\label{fig:4-corner}}
\end{figure}

Consider the 2D ``diamond shaped'' lattice of fig.\ref{fig:4-corner}
and divide it in four quadrants. The central site is denoted by 0.
In the lower-left quadrant $\boldsymbol{\mathsf{A}}$ introduce an
operator (see fig. \ref{fig:4-corner})
\[
\mathbf{A}_{\Bell,\Bell'}^{(N)}(u)={\displaystyle \delta_{\ell_{0},\ell_{0}^{\prime}}\sum_{\ell_{i}|i=\bullet\in\boldsymbol{\mathsf{A}}}\prod_{\square\in\boldsymbol{\mathsf{A}}}}W\left(\left.\begin{array}{cc}
\ell_{l} & \ell_{k}\\
\ell_{i} & \ell_{j}
\end{array}\right|u\right)
\]
where $\Bell=(\ell_{0},\ell_{1},\ell_{2},...,\ell_{N})$ and $\Bell'=(\ell_{0}^{\prime},\ell_{1}^{\prime},\ell_{2}^{\prime},...,\ell_{N}^{\prime})$
are vectors collecting all the variables along the two inner boundaries,
i.e. on the sites denoted by $\circ$ in fig. \ref{fig:4-corner}.
The sum is performed over all possible values of $\ell_{i}$ on internal
sites (signed by a black dot $\bullet$ in fig. \ref{fig:4-corner}).
The variables at the outer boundary sites are assigned fixed values
determining a boundary condition uniquely. The product is over all
tiles $\square$ of the quadrant. Notice the $\ell_{0}=\ell_{0}^{\prime}$
obvious constraint on the central site 0.

Analogously, define in the other quadrants the operators $\mathbf{B}_{\Bell',\Bell''}^{(N)}$,
$\mathbf{C}_{\Bell'',\Bell'''}^{(N)}$ and $\mathbf{D}_{\Bell''',\Bell}^{(N)}$.
Organizing the product of $W$'s to be performed diagonally (thus
the diamond shape of the lattice), a corner transfer matrix like $\mathbf{A}_{\Bell,\Bell'}^{(N)}$
may be expressed in terms of smaller corner transfer matrices $\mathbf{A}_{\Bell,\Bell'}^{(N-1)}$
and $\mathbf{A}_{\Bell,\Bell'}^{(N-2)}$ (defined for reduced quadrants).
This recursion relation allows, in principle, the iterative calculation
of the corner transfer matrix for any lattice quadrant of finite size.

The partition function can be expressed in terms of the four CTM operators
as
\[
Z^{(N)}=\mbox{tr}(\mathbf{A}^{(N)}\mathbf{B}^{(N)}\mathbf{C}^{(N)}\mathbf{D}^{(N)})
\]
There are other interesting quantities that can be computed form the
CTM method, like e.g. the height probabilities. Here we are interested
in the calculation of the reduced density matrix $\rho_{A}$ relative
to a dominion $A$ coinciding with the negative axis of the horiziontal
direction of the infinite lattice that results taking the thermodynamic
limit $N\to\infty$ of the present construction. Following \cite{CTMPeschel},
the unnormalised reduced density matrix on the subsystem $A$ can
be written as\footnote{Please notice the difference of the symbols $\varrho$ (unnormalised
density matrix) and $\rho$ (normalised density matrix)}
\[
\varrho_{A,\Bell\Bell'}=(\mathbf{ABCD})_{\Bell\Bell'}
\]
where $\mathbf{A}=\lim_{N\to\infty}\mathbf{A}^{(N)}$ and analogously
for \textbf{B}, \textbf{C}, \textbf{D}. Obviously $\mbox{tr}_{A}\varrho_{A}=\mathcal{Z}_{1}$
is the partition function in the thermodynamic limit 
\[
\mathcal{Z}_{1}=\lim_{N\to\infty}Z^{(N)}
\]
To normalise, one has to divide by the partition function
\[
\rho_{A,\Bell\Bell'}=\frac{(\mathbf{ABCD})_{\Bell\Bell'}}{\mathcal{Z}_{1}}
\]
For the Rényi Entanglement Entropy we need to compute $\mbox{tr}_{A}\rho_{A}^{n}$.
Defining the higher genus partition functions $\mbox{tr}_{A}\varrho_{A}^{n}=\mathcal{Z}_{n}$
one sees that
\[
\mbox{tr}_{A}\rho_{A}^{n}=\frac{\mathcal{Z}_{n}}{\mathcal{Z}_{1}^{n}}
\]
and the expression for the Rényi Entropy is
\[
S_{A}^{(n)}=\frac{1}{1-n}\log\frac{\mathcal{Z}_{n}}{\mathcal{Z}_{1}^{n}}
\]

As shown by Baxter \cite{Baxter-book}, the Yang-Baxter equation that
has to be satisfied by the Boltzamnn weights of any integrable model,
implies that the four CTM operators, for $N\to\infty$, commute each
other for different values of their parameters and their eigenvalues,
up to a common factor, accommodate into a diagonal matrix of the general
form
\[
\varrho_{A,\Bell\Bell'}^{\mathrm{diag}}=R(\ell_{0})T(\Bell)\delta_{\Bell,\Bell'}
\]
The diagonalisation can be performed along the lines illustrated in
\cite{Baxter-book}. In particular, for FB RSOS models it has been
performed in the original paper \cite{FB}. It is simplified if one
introduces new variables $x$, $y$ that in regime III are defined
as
\begin{equation}
y=e^{\frac{4\pi^{2}}{\log p}}\qquad,\qquad x=e^{\frac{4\pi^{2}}{\log p}\frac{s}{r}}=y^{\frac{s}{r}}\label{X=000026Y}
\end{equation}

Define
\[
\Phi(\Bell)=\sum_{k=1}^{N}k\gamma(\ell_{k},\ell_{k+1},\ell_{k+2})
\]
where\footnote{$\left\lfloor x\right\rfloor $ denotes the integer part of $x$.}
\[
\gamma(\ell,\ell\pm1,\ell)=\mp\left\lfloor \frac{\ell s}{r}\right\rfloor \qquad,\qquad\gamma(\ell\pm1,\ell,\ell\mp1)=\frac{1}{2}
\]
In regime III we have 
\[
\Phi(\Bell)=\sum_{k=1}^{N}\left\{ k\frac{|\ell_{k}-\ell_{k+2}|}{4}+(\ell_{k}-\ell_{k+1})\left\lfloor \frac{\ell_{k}s}{r}\right\rfloor \right\} 
\]
The second term of the summand is absent in the $s=1$ case simply
because $\left\lfloor \ell_{k}/r\right\rfloor $ is always 0. Consider
now a system with size $2N+3$. The unormalised reduced density matrix
in regime III, as computed in \cite{FB}, is given by 
\[
\varrho_{A,\Bell\Bell'}^{(N)}=\left(\mathbf{ABCD}\right)_{\Bell\Bell'}^{(N)}=E\left(x^{\ell_{0}},y\right)x^{2\Phi(\Bell)}\delta_{\Bell\Bell'}
\]
where $\ell_{0}$ is the central height and $\Bell=(\ell_{0},\dots,\ell_{N+1})$
is a vector of all local heights from the central to the boundary
$\ell_{N+1}$. Here $E(\cdot,\cdot)$ is the modular form defined
in (\ref{eq:E}).

\section{Rényi Entropy in FB RSOS models\label{EERSOS}}

As disccussed in section \ref{sec:R=0000E9nyi-Entanglement-Entropies},
to compute the Rényi Entropy we need to consider not only $\varrho_{A}$
but also its $n$-th powers $\varrho_{A}^{n}$.

Let us set the boundary condition to $\ell_{N}=b$ and $\ell_{N+1}=c=b\pm1$.
The $n$-th power of the reduced density matrix is given by 
\[
(\varrho_{A}^{(N)})_{\boldsymbol{\ell}\boldsymbol{\ell}'}^{n}=E\left(x^{\ell_{0}},y\right)^{n}x^{2n\Phi(\boldsymbol{\ell})}\delta_{\boldsymbol{\ell}\boldsymbol{\ell}'}
\]
The trace can be performed summing over all allowed configurations
$\boldsymbol{\ell}$
\begin{eqnarray}
Z_{n}^{(N)} & = & \sum_{\boldsymbol{\ell}}E\left(x^{\ell_{0}},y\right)^{n}x^{2n\Phi(\boldsymbol{\ell})}\nonumber \\
 & \equiv & \ssum{a=1}{r-1}E\left(x^{a},y\right)^{n}D_{N}\left(a,b,c;x^{2n}\right)\label{eq:Zn}
\end{eqnarray}
 where $D_{N}(a,b,c;q)=\sum\limits _{\ell_{1}\cdots\ell_{N-1}}q^{\Phi(\boldsymbol{\ell})}$
with $\ell_{0}=a$, $\ell_{N}=b$ and $\ell_{N+1}=c$. The $\sum^{*}$
symbol means that the sum is restricted to even or odd only values
of $a$ accordingly to the parity of the boundary conditions. In particular
notice the the central height $a$ and the boundary height $b$ have
to be compatible with the requirement $|\ell_{i}-\ell_{i+1}|=1$.

The limit $N\rightarrow\infty$ can be performed, keeping track of
the boundary condition $\ell_{N}\lessgtr\ell_{N+1}$: 
\[
\lim_{N\rightarrow\infty}q^{-\frac{kN}{2}}D_{N}(a,b,b+1;q)=\frac{1}{(q)_{\infty}}q^{\frac{b(b-1)}{4}-\frac{(k-1)b}{2}}F(a,b-k;q)
\]
with $k=\lfloor\frac{s(b+1)}{r}\rfloor=\lfloor\frac{s\ell_{N+1}}{r}\rfloor$
if $\ell_{N}<\ell_{N+1}$ or 
\[
\lim_{N\rightarrow\infty}q^{\frac{kN}{2}}D_{N}(a,b+1,b;q)=\frac{1}{(q)_{\infty}}q^{\frac{b(b+1)}{4}-\frac{k(b+1)}{2}}F(a,b-k;q)
\]
with $k=\lfloor\frac{sb}{r}\rfloor=\lfloor\frac{s\ell_{N+1}}{r}\rfloor$
if $ $$\ell_{N}>\ell_{N+1}$.

The function $F$ is defined as 
\begin{eqnarray}
F(a,d;q) & = & q^{\frac{a(a-1)}{4}-\frac{ad}{2}}\left[E\left(-q^{rd+(r-a)(r-s)},q^{2r(r-s)}\right)-q^{ad}E\left(-q^{rd+(r-a)(r+s)},q^{2r(r-s)}\right)\right]\nonumber \\
 & = & q^{\frac{a(a-1)}{4}-\frac{ad}{2}}\frac{(q)_{\infty}}{q^{-\frac{c}{24}+\Delta_{da}}}\chi_{da}(q)\label{eq:F}
\end{eqnarray}
where the Virasoro characters $\chi_{da}(q)$ (see eq.(\ref{eq:character-E}))
are taken here for $m=r-s$ and $m'=r$. Thus
\[
\lim_{N\rightarrow\infty}q^{\mp\frac{kN}{2}}D_{N}(a;q)_{b,d=b-k}=q^{f_{\mp}(b,d)}q^{\frac{a(a-1)}{4}-\frac{ad}{2}+\frac{c}{24}-\Delta_{da}}\chi_{da}(q)
\]
where 
\begin{eqnarray}
f_{+}(b,d) & = & \frac{b(b-1)}{4}-\frac{b(k-1)}{2}\nonumber \\
f_{-}(b,d) & = & \frac{b(b+1)}{4}-\frac{k(b+1)}{2}\label{eq:f+-}
\end{eqnarray}
do not depend on $a$.

Since the physical (normalised) density operator is given by 
\[
\rho_{A}\equiv\frac{\varrho_{A}}{\mbox{Tr}_{A}\varrho_{A}}=\frac{\mathbf{ABCD}}{\mbox{Tr}(\mathbf{ABCD})}
\]
the quantity in which we are interested is 
\[
\mbox{tr}_{A}{\rho_{A}}^{n}=\frac{\mathcal{Z}_{n}}{\mathcal{Z}_{1}^{n}}=\frac{\hat{\mathcal{Z}}_{n}}{\hat{\mathcal{Z}}_{1}^{n}}
\]
where
\begin{equation}
\hat{\mathcal{Z}}_{n}={\displaystyle \ssum{a=1}{r-1}}E\left(x^{a},y\right)^{n}x^{2n\left(\frac{a(a-1)}{4}-\frac{ad}{2}-\Delta_{da}\right)}\chi_{da}\left(x^{2n}\right)\label{Zhat}
\end{equation}
 since the factor $\left(f_{\mp}+\frac{c}{24}\right)^{n}$ appears
in both the numerator and the denominator and cancels out.

Notice that the critical point is reached for $p\rightarrow0$, or
$x\rightarrow1$. In order to catch the critical behaviour we transform
(\ref{Zhat}) into a new expression which is more suitable for expansion
near $p=0$. First of all we can use the relations (\ref{eq:E_theta1})
and (\ref{eq:theta1_modular}) for the function $E$ 
\[
\mathcal{W}_{n}=\ssum{a=1}{r-1}\vartheta_{1}\left(\frac{\pi as}{r},\sqrt{p}\right)^{n}\chi_{da}\left(x^{2n}\right)
\]
where, again, $\mathcal{W}_{n}$ is proportional to $\hat{\mathcal{Z}_{n}}$
and a common factor with $\hat{\mathcal{Z}}_{1}^{n}$ has been dropped
so that 
\[
\mbox{tr}_{A}{\rho_{A}}^{n}=\frac{\mathcal{Z}_{n}}{\mathcal{Z}_{1}^{n}}=\frac{\hat{\mathcal{Z}}_{n}}{\hat{\mathcal{Z}}_{1}^{n}}=\frac{\mathcal{W}_{n}}{\mathcal{W}_{1}^{n}}
\]

Furthermore we can perform a modular $\hat{\mathcal{S}}$ transformation
(\ref{eq:modular}) on the conformal character 
\[
\chi_{a,a'}\left(\tilde{q}\right)=\sum_{(b,b')\in\mathcal{J}}\mathcal{S}_{a,a'}^{b,b'}\chi_{b,b'}(q)
\]
Since $x^{2}=e^{\frac{8\pi^{2}}{\log p}\frac{s}{r}}=e^{2\pi i\left(-i\frac{4\pi}{\log p}\frac{s}{r}\right)}$,
$ $its modular $\hat{\mathcal{S}}$-transform is $\omega\equiv\widetilde{\left(x^{2}\right)}=e^{\frac{\log p}{2}\frac{r}{s}}=p^{\frac{r}{2s}}$
(recall that, in regime III, $p>0$). Similarly, $\widetilde{(x^{2n})}=(p^{\frac{r}{2s}})^{\frac{1}{n}}=\omega^{1/n}$.
We have 
\[
{\mathcal{W}}_{n}=\ssum{a=1}{r-1}\sum_{d'a'}\vartheta_{1}\left(\frac{\pi as}{r},\sqrt{p}\right)^{n}\mathcal{S}_{da}^{d'a'}\chi_{d'a'}\left(\omega^{\frac{1}{n}}\right)
\]
which is suitable for a $p\rightarrow0$ expansion.

Denoting $\mathfrak{h}=(d',a')\in\mathcal{J}$ as an index which spans
the Kac table we have 
\begin{eqnarray}
\mathcal{W}_{n} & = & \sum_{\mathfrak{h}\in\mathcal{J}}\chi_{\mathfrak{h}}\left(\omega^{\frac{1}{n}}\right)f_{\mathfrak{h}}\left(n,p\right)\nonumber \\
 & = & \chi_{\mathrm{min}}\left(\omega^{\frac{1}{n}}\right)f_{\mathrm{min}}\left(n,p\right)\left(1+\sum_{\mathfrak{h}\neq\mathrm{min}}\frac{\chi_{\mathfrak{h}}\left(\omega^{\frac{1}{n}}\right)}{\chi_{\mathrm{min}}\left(\omega^{\frac{1}{n}}\right)}\frac{f_{\mathfrak{h}}\left(n,p\right)}{f_{\mathrm{min}}\left(n,p\right)}\right)
\end{eqnarray}
where 
\[
f_{\mathfrak{h}}\left(n,p\right)=\ssum{a=1}{r-1}\vartheta_{1}\left(\frac{\pi as}{r},\sqrt{p}\right)^{n}\mathcal{S}_{da}^{d'a'}
\]
and $\mathfrak{h}=\mathrm{min}$ refers to the primary field with
the lowest conformal dimension $\Delta_{\mathrm{min}}$.

Taking the logarithm, we have 
\begin{eqnarray}
\log\frac{\mathcal{W}_{n}}{{\mathcal{W}_{1}}^{n}} & = & \log\frac{\chi_{\mathrm{min}}\left(\omega^{\frac{1}{n}}\right)}{\chi_{\mathrm{min}}\left(\omega\right)^{n}}+\log\frac{f_{\mathrm{min}}\left(n,p\right)}{f_{\mathrm{min}}\left(1,p\right)^{n}}\nonumber \\
 & + & \log\left(1+\sum_{\mathfrak{h}\neq\mathrm{min}}\frac{\chi_{\mathfrak{h}}\left(\omega^{\frac{1}{n}}\right)}{\chi_{\mathrm{min}}\left(\omega^{\frac{1}{n}}\right)}\frac{f_{\mathfrak{h}}\left(n,p\right)}{f_{\mathrm{min}}\left(n,p\right)}\right)\nonumber \\
 & - & n\log\left(1+\sum_{\mathfrak{h}\neq\mathrm{min}}\frac{\chi_{\mathfrak{h}}\left(\omega\right)}{\chi_{\mathrm{min}}\left(\omega\right)}\frac{f_{\mathfrak{h}}\left(1,p\right)}{f_{\mathrm{min}}\left(1,p\right)}\right)
\end{eqnarray}
Expanding $\chi_{\mathrm{min}}$ and $f_{\mathrm{min}}$ of the first
two terms of the equation above near $p,\omega=0$ we obtain the leading
scaling and the constant coefficient for the Rényi Entropy 
\begin{eqnarray}
S_{A}^{(n)} & = & -\frac{\ceff}{24}\frac{n+1}{n}\log\omega+\tilde{A}^{(n)}\nonumber \\
 & + & \frac{1}{1-n}\log\left(1+\sum_{\mathfrak{h}\neq\mathrm{min}}\frac{\chi_{\mathfrak{h}}\left(\omega^{\frac{1}{n}}\right)}{\chi_{\mathrm{min}}\left(\omega^{\frac{1}{n}}\right)}\frac{f_{\mathfrak{h}}\left(n,p\right)}{f_{\mathrm{min}}\left(n,p\right)}\right)\label{eq:Renyi_final_1}\\
 & - & \frac{n}{1-n}\log\left(1+\sum_{\mathfrak{h}\neq\mathrm{min}}\frac{\chi_{\mathfrak{h}}\left(\omega\right)}{\chi_{\mathrm{min}}\left(\omega\right)}\frac{f_{\mathfrak{h}}\left(1,p\right)}{f_{\mathrm{min}}\left(1,p\right)}\right)
\end{eqnarray}
The constant $\tilde{A}^{(n)}$ is given by 
\begin{equation}
\tilde{A}^{(n)}=\frac{1}{1-n}\log\frac{\sideset{}{^{*}}\sum\limits _{{a=1}}^{{r-1}}\mathcal{S}_{da}^{\mathrm{min}}\sin^{n}\frac{\pi as}{r}}{\left(\sideset{}{^{*}}\sum\limits _{{a=1}}^{{r-1}}\mathcal{S}_{da}^{\mathrm{min}}\sin\frac{\pi as}{r}\right)^{n}}\label{AA}
\end{equation}
which is well defined in the $n\rightarrow1$ limit.

\section{``Unusual'' corrections\label{sec:Unusual-corrections}}

The next step is the evaluation of power law corrections to the logarithmic
scaling of the Rényi Entropy. First, notice that $\chi_{\mathfrak{h}}\ll\chi_{\mathrm{min}}$
for $\omega\rightarrow0$. In this regime the argument of the logarithms
in the second and third line of (\ref{eq:Renyi_final_1}) is close
to $1$ and then it can be Taylor expanded. Taking into account only
the most relevant contribution and assuming, to fix ideas, that $n>1$,
we have 
\[
S_{A}^{(n)}=-\frac{\ceff}{24}\frac{n+1}{n}\log\omega+\tilde{A}^{(n)}+\sum_{\mathfrak{h}\neq\mathrm{min}}\tilde{B}_{\mathfrak{h}}^{(n)}\omega^{\frac{\Delta_{\mathfrak{h}}-\Delta_{\mathrm{min}}}{n}}+\cdots
\]
with 
\begin{equation}
\tilde{B}_{\mathfrak{h}}^{(n)}=\frac{1}{1-n}\frac{\sum\limits _{a}\sin^{n}\frac{\pi as}{r}\mathcal{S}_{da}^{\mathfrak{h}}}{\sum\limits _{a}\sin^{n}\frac{\pi as}{r}\mathcal{S}_{da}^{\mathrm{min}}}\label{BB}
\end{equation}
and the most relevant contribution is given by the second smallest
conformal dimension among those appearing in the expansion of \ref{eq:Renyi_final_1},
denoted $\Delta_{1}$ in the following. Taking into account only this
correction we have 
\[
S_{A}^{(n)}=-\frac{\ceff}{24}\frac{n+1}{n}\log\omega+\tilde{A}^{(n)}+\tilde{B}_{1}^{(n)}\omega^{\frac{\Delta_{1}-\Delta_{\mathrm{min}}}{n}}+\cdots
\]
In the case $n<1$ one can proceed similarly, but now the first correction
will be
\[
S_{A}^{(n)}=-\frac{\ceff}{24}\frac{n+1}{n}\log\omega+\tilde{A}^{(n)}+\tilde{B}_{1}^{\prime(n)}\omega^{\Delta_{1}-\Delta_{\mathrm{min}}}+\cdots
\]
with
\[
\tilde{B}_{\mathfrak{h}}^{\prime(n)}=\frac{n}{1-n}\frac{\sum\limits _{a}\sin\frac{\pi as}{r}\mathcal{S}_{da}^{\mathfrak{h}}}{\sum\limits _{a}\sin\frac{\pi as}{r}\mathcal{S}_{da}^{\mathrm{min}}}=n\tilde{B}_{\mathfrak{h}}^{(1)}
\]
In order to interpret these results from a physical point of view,
i.e. to have an expression for the entropy in terms of the correlation
length (or in term of the mass), we need a relation between the parameter
$\omega=p^{\frac{r}{2s}}$ and the correlation length $\xi$. In other
words, we need to know the critical exponent $\nu$ in regime III:
\[
m\;\;\;=\;\;\;\xi^{-1}\sim p^{\nu}
\]

Using perturbative CFT it is possible to evaluate the critical exponent
$\nu$. In the continuum limit, the model is described by the perturbed
CFT: 
\[
S=S_{\mathrm{CFT}}+t\int d^{2}x\;\Phi_{1,3}(x)
\]
where $|t|=p^{2}$ measures the departure from criticality. A simple
dimensional analysis of this action tells us that $\nu=\frac{1}{2(1-\Delta_{1,3})}=\frac{r}{4s}$
for the minimal model $\mathcal{M}_{r-s,r}$ at the transition between
regime III and IV.

This value for the critical exponent $\nu$ can also be extracted
extending previous results \cite{OBrian} to the Forrester Baxter
models. Taking into account the necessary modifications, the calculation
can be carried out along the same lines of \cite{OBrian} to get 
\begin{equation}
e^{-\frac{1}{\xi}}=k'(p^{\nu})\label{eq:xi-exact}
\end{equation}
where $k'(q)$ is the conjugate elliptic modulus for the elliptic
nome $q$ 
\[
k'(q)=\prod_{\ell=1}^{\infty}\left(\frac{1-q^{2\ell-1}}{1+q^{2\ell-1}}\right)^{4}
\]
Expanding (\ref{eq:xi-exact}) around $p=0$ we get 
\begin{equation}
\xi^{-1}=8p^{\nu}+\frac{32}{3}p^{3\nu}+\frac{48}{5}p^{5\nu}+\frac{64}{7}p^{7\nu}+O(p^{9\nu})\label{xi=000026p}
\end{equation}
in perfect agreement with the perturbative CFT prediction.

Using (\ref{X=000026Y}) and (\ref{xi=000026p}) we have $\omega=(8\xi)^{-2}+\cdots$
which gives 
\[
S_{A}^{(n)}=\frac{\ceff}{12}\frac{n+1}{n}\log\xi+A^{(n)}+B_{1}^{(n)}\xi^{-\frac{2}{n}(\Delta_{1}-\Delta_{\mathrm{min}})}+nB_{1}^{(1)}\xi^{-2(\Delta_{1}-\Delta_{\min})}+...
\]
where constants $A^{(n)}$ and $B_{1}^{(n)}$ are just trivial rescaling
of $\tilde{A}^{(n)}$ and $\tilde{B}_{1}^{(n)}$: 
\begin{eqnarray}
A^{(n)} & = & \tilde{A}^{(n)}+\frac{\ceff}{4}\frac{n+1}{n}\log2\label{Renyi_final_4}\\
B_{1}^{(n)} & = & 8^{-\frac{2}{n}(\Delta_{1}-\Delta_{\mathrm{min}})}\tilde{B}_{1}^{(n)}
\end{eqnarray}
From the equation (\ref{Renyi_final_4}), it is immediately possible
to read the correction to the logarithmic scaling. For unitary theories
the correction is expected to scale as $\xi^{-\frac{2\Delta}{n}}$
where $\Delta$ is the conformal dimension of the field related to
the correction \cite{EEABF}. In the non unitary case this term is
affected by the fact that the ground state is no more the conformal
vacuum $|0\rangle$ but another state $|\mathrm{min}\rangle=\Phi_{\mathrm{min}}(0,0)|0\rangle$.
The formula $\Delta_{1}-\Delta_{\mathrm{min}}$ shows that the presence
of a non trivial ground state modifies the scaling of the correction.
It is known \cite{NostroLondra} that this feature affects also the
logarithmic scaling of the entropy in non-unitary models, where the
central charge $c$ is replaced by the effective one $\ceff$. Similarly,
it is not surprising that the conformal dimensions are replaced by
some sort of ``effective'' dimensions $\Delta-\Delta_{\min}$. Notice
that the unitary case can be immediately recovered in (\ref{Renyi_final_4})
by setting $\Delta_{\mathrm{min}}=0$.

\section{A comment on off-critical Logarithmic Minimal Models\label{sec:Log-CFT}}

An interesting feature of Forrester Baxter RSOS models is that, for
a particular choice of parameters $r$ and $s$, they provide a lattice
realisation of Logarithmic Minimal models \cite{Pearce-log-minimal}
and their off-critical thermal perturbations \cite{PearceSeaton}.
In particular it has been shown that in the limit $r,s\to\infty$,
keeping the ratio $r/s$ fixed to some rational number $R/S$, the
underlying model becomes the so called logarithmic minimal model $\mathcal{LM}_{R-S,R}$
and its off-critical thermal perturbation $\mathcal{LM}_{R-S,R}+t\Phi_{1,3}$
.

Taking such a limit for the entropy, the result is not affected from
the functional point of view: it maintains the same structure and
the effective central charge 
\[
\ceff=\lim_{r,s\to\infty}\left(1-\frac{6}{r(r-s)}\right)
\]
is identically 1 for any choice of the ratio $R/S$ 
\begin{equation}
S_{A}^{(n)}=\frac{1}{12}\frac{n+1}{n}\log\xi+\bar{A}^{(n)}+\bar{B}_{1}^{(n)}\xi^{-\frac{2}{n}(\Delta_{1}-\Delta_{\mathrm{min}})}+n\bar{B}_{1}^{(1)}\xi^{-2(\Delta_{1}-\Delta_{\mathrm{min}})}+\cdots\label{eq:Renyi-log}
\end{equation}
Here $\bar{B}_{1}^{(n)}={\displaystyle \lim_{r,s\to\infty}}B_{1}^{(n)}$.
When taking the limit $r,s\to\infty$ the $\mathcal{S}$ modular matrix
becomes ill defined because the prefactor $\sqrt{\frac{2}{r(r-s)}}=\frac{1}{r}\sqrt{\frac{2}{1-\frac{s}{r}}}$
tends to zero in this limit. While this feature does not affect the
limit of the coefficent $B_{1}^{(n)}$ (\ref{BB}) -- as $\mathcal{S}$
appears with the same power both at the numerator and at the denominator
-- it implies some extra attention for the limit of the constant $A^{(n)}$
(\ref{AA}). For this reason, we need to modify the definition of
the Rényi Entropy for Logarithmic models multiplying the partition
function by $r$ in a sort of renormalisation procedure
\[
\bar{A}^{(n)}=\lim_{r,s\to\infty}\left(A^{(n)}-\log\frac{1}{r}\right)
\]
This multiplication can be seen in the same spirit of defyning generalised
order parameters in \cite{PearceSeaton}, while a better understanding
of this renormalistion of the Entropy is still missing and goes far
beyond the aim of this work. In any case, it only affects the numerical
value of the non-universal constant $\bar{A}^{(n)}$ and not the functional
shape of the $\xi$ dependence of the entropy.

This $\xi$ dependence result (\ref{eq:Renyi-log}) disagrees with
the general prediction for the Rényi Entropy $l$ dependence in logarithmic
CFT \cite{NostroLondra}, where a double logarithmic (log log) term
is expected. The difference is due to the fact that the logarithmic
feature of the system is related to the presence of non diagonalisable
Jordan blocks in the Hamiltonian. It has been noticed in \cite{PearceSeaton}
that such non diagonalisability feature disappears as soon as the
system is perturbed thermally out of the critical point. Thus the
so called off-critical logarithmic minimal models are not really logarithmic
and thus we should not expect double log term in the computation of
the entropy. In the case of logarithmic minimal models, then, we expect
that, while the dependence in the subsystem size $l$ at criticality
rescales with a double logarithmic term, this is absent in the rescaling
off-criticality in $\xi$ while approaching the logarithmic critical
point. The two behaviours in this case are different, while in the
usual minimal models (unitary or not) they show a similar (single
log rescaling) pattern.

\section{Conclusions and outlook\label{sec:Conclusions-and-outlook}}

We have computed the Entanglement Rényi Entropy for Forrester-Baxter
RSOS models in the regime III, a set of lattice systems whose continuum
limit is described by non-unitary Minimal Conformal Models perturbed
by their $\Phi_{1,3}$ operator. Our evaluation focused on the scaling
with the correlation length $\xi$ of the Entropy for a infinite,
bipartite, quantum system. The computation of Entanglement Entropy
in non unitary models has been recently addressed in \cite{NostroLondra}
and \cite{EENUQFT} where a logarithmic scaling with the size $l$
of the subsystem has been found. In particular it has been shown in
\cite{NostroLondra} that the coefficient of the logarithm is given
by the \textit{effective} central charge $\ceff$. For unitary theories
the computation of the Entropy for a finite critical system ($l<\infty,\;m^{-1}=\xi=\infty$)
can be easily translated \cite{CardyCalabrese} in the infinite subsystem
with a small but non zero mass gap ($l=\infty,\;m^{-1}=\xi<\infty$)
using arguments similar to the Zamolodchikov c-theorem. In the non
unitary case such a translation cannot be performed \textit{a priori},
since some assumptions about the positivity of certain correlation
functions are no more valid in the non unitary case. For this reason,
the scaling we have found, where the proper length $l$ is replaced
by the correlation length $\xi$ gives a strong hint about the renormalisation
flux properties in the non unitary case. A proof of a sort of $\ceff$-theorem
in the non unitary case goes far beyond the aim of this paper. Nevertheless,
our result supports evidence, in a non trivial set of non unitary
models, that the $\xi$ scaling follows the same logarithmic law of
the $l$ scaling, exactly like in the unitary models.

We have also studied the power law corrections to the logarithmic
scale of the Entropy. The interesting part is not in the coefficients
of the expansion, that are non universal, but in the exponent of the
power of $\xi$. In the unitary case the expansion organises into
many series expansions in terms of powers of $\xi^{\Delta_{\mathfrak{h}}}$
or $\xi^{\Delta_{\mathfrak{h}}/n}$, with $\mathfrak{h}$ labelling
the various conformal primary fields. In the non unitary case we find
that the series expansions are in terms of $\xi^{\Delta_{\mathfrak{h}}-\Delta_{\min}}$
or $\xi^{(\Delta_{\mathfrak{h}}-\Delta_{\min})/n}$. In other words,
also the conformal dimensions here take an ``effective'' value shifted
by $\Delta_{\min}$, exactly like the central charge is replaced by
$\ceff$.

We have also briefly considered the limiting case of the off-critical
Logarithmic Minimal Models. Even if a double logarithmic term has
been expected from the literature \cite{NostroLondra}, where the
$l$ rescaling has been considered, we did not find here such a term:
the scaling in $\xi$ behaves exactly as in the non-logarithmic case.
This discrepancy is due to the fact that the perturbation outside
the critical point destroys the Jordan blocks responsible for the
creation of the logarithmic features, as already pointed out in \cite{PearceSeaton}.

It their seminal work on the FB RSOS model \cite{FB}, Forrester and
Baxter classified many regimes, while we have restricted our analysis
to regime III only. While it is known \cite{Riggs,Nakanishi} that
the regime III is a lattice realisation of the perturbed Minimal Models
$\mathcal{M}_{r-s,r}+\Phi_{1,3}$, the physical interpretation of
the other regimes has still to be explored. In the unitary ABF RSOS
model \cite{ABF}, the number of regimes is smaller (just four, compared
to ten in the FB case) and two kind of universality classes have been
identified. The critical line between the Regime III and IV can be
described by the Unitary Minimal Models $\mathcal{M}_{r-1,r}$, while
a unitary parafermionic CFT $\mathbb{Z}_{r-2}$ underlies the critical
line between regimes I and II. It would be very interesting to try
to understand if some critical line of the FB RSOS model can be classified
using, maybe, the non unitary parafermionc CFT introduced in \cite{Ahn}
or if they obey instead some other classification scheme. The evaluation
of Entanglement Entropy has been demonstrated to be a powerful tool
for the identification of universality classes and it could be a valid
help also in the study of the unknown regimes of the FB RSOS model.

Of course, further generalizations of these results to the lattice
realisations of higher coset conformal filed theories perturbed by
their $\Phi_{1,1,3}$ opertors, or even for models based on cosets
of higher algebras, like $W_{n}$-algebra based series, including
Pasquier like generalizations and their dilute versions, etc... can
all be the subject of future investigation. As non unitary coset models
as critical points of fusions of FB RSOS models are under construction
right in these days \cite{Pearce-nuovo}, the exploration of Rényi
Entanglement Entropies for this set of models is viable.

The bipartite Entanglement Entropy can give a lot of useful information,
but a more complete understanding of entanglement needs also the knowledge
of the dependence on the size of the system, as well as its behaviour
in finite intervals. We know how to deal with these problems in CFT
\cite{CardyCalabrese} and, with a form factor expansion, also in
off-critical integrable quantum field theories \cite{EEQFT}. However,
in integrable spin chains, a useable procedure to compute entanglement
entropies on a finite interval is still lacking in general. This would
be a serious progress in our understanding of entanglement in one-dimensional
systems.

Also, the interest in entanglement measurements with more than two
subsystems introduces new quantites to be evaluated, like negativity
\cite{negativity1,negativity2}. To find a way to compute these quantites
in a consistent lattice integrable approach is a challenge for future
research.

\subsection*{Acknowledgments}

We are endebted to O. A. Castro-Alvaredo and B. Doyon for very useful
discussions. In particular, we thank the fruitful and deep exchange
of ideas with P. A. Pearce, as well as his continuous and encouraging
interest in this work. We acknowledge E. Ercolessi for the advice
and the deep discussions during the elaboration of the M.Sc. thesis
of one of us (D.B.), where many of the results presented here were
\emph{in nuce} obtained. F.R. thanks INFN, and in particular its grant
GAST, for partial financial support. D.B. thanks City University London
for his Doctoral Scholarship and the University of Bologna for the
warm hospitality when part of this work has been done.

\end{document}